\documentclass[a4paper]{spie}  

\usepackage{amsmath,amsfonts,amssymb}
\usepackage{graphicx}
\usepackage[colorlinks=true, allcolors=blue]{hyperref}

\title{Study of the HV power supply modules for the CUbesat Solar
Polarimeter (CUSP)}

\author[a]{Alessandro Lacerenza}
\author[a]{Alda Rubini}
\author[a,b]{Andrea Alimenti}
\author[a]{Sergio Fabiani}
\author[a]{Ettore Del Monte}
\author[g,n]{Riccardo Campana}
\author[h]{Mauro~Centrone}
\author[a]{Enrico Costa}
\author[a]{Nicolas De Angelis}
\author[g]{Giovanni~De~Cesare}
\author[a]{Sergio~Di~Cosimo}
\author[a]{Giuseppe Di Persio}
\author[a]{Abhay Kumar}
\author[a]{Pasqualino~Loffredo}
\author[a,c]{Giovanni Lombardi}
\author[l]{Gabriele Minervini}
\author[a]{Fabio~Muleri}
\author[m]{Paolo Romano}
\author[a]{Emanuele Scalise}
\author[a,b]{Enrico Silva}
\author[a]{Paolo~Soffitta}
\author[d]{Davide~Albanesi}
\author[e]{Ilaria Baffo}
\author[f]{Daniele Brienza}
\author[d]{Valerio Campamaggiore}
\author[i]{Giovanni~Cucinella}
\author[j]{Andrea Curatolo}
\author[d]{Giulia de Iulis}
\author[d]{Andrea Del Re}
\author[i]{Vito Di Bari}
\author[i]{Simone~Di~Filippo}
\author[f]{Immacolata Donnarumma}
\author[e]{Pierluigi Fanelli}
\author[k]{Nicolas~Gagliardi}
\author[d]{Paolo~Leonetti}
\author[f]{Matteo Mergè}
\author[j,k]{Dario~Modenini}
\author[i]{Andrea Negri}
\author[j]{Daniele~Pecorella}
\author[i]{Massimo~Perelli}
\author[k]{Alice~Ponti}
\author[d]{Francesca Sbop}
\author[j,k]{Paolo~Tortora}
\author[f]{Alessandro Turchi}
\author[f]{Valerio~Vagelli}
\author[f]{Emanuele Zaccagnino}
\author[d]{Alessandro Zambardi}
\author[e]{Costantino Zazza}

\affil[a]{INAF-IAPS\\ via del Fosso del Cavaliere 100, 00133, Rome, Italy}
\affil[b]{Department of Industrial, Electronic and Mechanical Engineering, "Roma Tre" University, via V. Volterra 62, 00146, Rome, Italy}
\affil[c]{Department of Enterprise Engineering "Mario Lucenti”, University of Rome "Tor Vergata", Via Cracovia 50, 00133, Rome, Italy}
\affil[d]{DEDA Connect s.r.l.\\ via Vincenzo Lamaro 51, 00173 Rome, Italy}
\affil[e]{DEIM, University of "La Tuscia", Largo dell’Università, 01100, Viterbo, Italy}
\affil[f]{ASI, via del Politecnico snc\\ 00133, Rome, Italy}
\affil[g]{INAF-OAS Bologna\\ via Gobetti 93/3, 40129, Bologna, Italy}
\affil[h]{INAF-OAR\\ via Frascati 33, 00040, Monte Porzio Catone, Italy}
\affil[i]{IMT s.r.l.\\ via Carlo Bartolomeo Piazza 30, 00161, Rome, Italy}
\affil[j]{Department of Industrial Engineering - Alma Mater Studiorum Università di Bologna - Via Montaspro 97, 47121 Forlì, Italy}
\affil[k]{Interdepartmental Centre for Industrial Aerospace Research - Alma Mater Studiorum Università di Bologna -  Via Carnaccini 12, 47121 Forlì, Italy}
\affil[l]{INAF-Headquarters\\ viale del Parco Mellini 84, 00136, Rome, Italy}
\affil[m]{INAF-OACT\\ Via S. Sofia 78, 95123, Catania, Italy}
\affil[n]{INFN Sezione di Bologna, viale Berti Pichat 6/2,  40127, Bologna, Italia}

\authorinfo{Further author information: (Send correspondence to A.L.)\\A.L.: E-mail: alessandro.lacerenza@inaf.it}

\begin{document} 
\maketitle

\begin{abstract}
The CUbesat Solar Polarimeter (CUSP) project is a CubeSat mission orbiting the Earth aimed to measure the linear polarization of solar flares in the hard X-ray band by means of a Compton scattering polarimeter. CUSP will allow to study the magnetic reconnection and particle acceleration in the flaring magnetic structures of our star. CUSP is a project in the framework of the Alcor Program of the Italian Space Agency aimed to develop new CubeSat missions. CUSP undergoing the Phase B started in December 2024 that will last for 12 month. The Compton polarimeter of the CUSP payload performs coincidence measurements between plastic scintilaltors and GaGG(Ce) crystals to derive the polarization of X-rays. These sensors are readout by Multi Anode Photomultiplier Tubes (MAPMTs) and Avalanche Photodiodes (APDs) respectively. Both sensors need an HV power supply  up to -1~kV (for the MAPMT) and +500~V (for the APD). We tested precision regulated High Voltage DC/DC Converters by HVM Technology Inc. with Sub-Miniature Case Size ($0.85''\times0.85''\times0.60''$) of the SMHV series. These modules are compact and suited for CubeSat missions.   
\end{abstract}


\section{INTRODUCTION}
\label{sec:intro}  

The CubeSat Solar Polarimeter (CUSP) is a CubeSat mission designed to study the linear polarization of solar flares in the hard X-ray energy range~\cite{papaioannou2016solar}. The polarization measurement is performed using a dual-phase Compton polarimeter~\cite{fabiani2014astronomical}, which relies on the analysis of X-ray photons that undergo Compton scattering in plastic scintillators, followed by absorption in surrounding GaGG(Gd$_3$Al$_2$Ga$_3$O$_{12}$) crystals~\cite{fabiani2022cusp,fabiani2024cubesat}. The GaGG scintillators are arranged all around the plastic scatterer, forming an efficient geometry for capturing scattered photons and maximizing the instrument’s sensitivity to polarization. The scattering events are detected by Multi-Anode Photomultiplier Tubes (MAPMTs), while the absorbing ones by Avalanche Photodiodes (APDs). 

In this work, we focus on the problem of gain stability in Avalanche Photodiodes (APDs) \cite{alimenti2024design,alimenti2024development}.  The CubeSat will not implement any active thermal control system to maintain a stable internal temperature.  Consequently, the payload will experience temperature variations within the expected range of $-20,^{\circ}\mathrm{C}$ to $+60,^{\circ}\mathrm{C}$, typical for CubeSats \cite{piedra2019thermal}.  Since it is well known that the APD gain, $G_{\mathrm{APD}}$, is strongly dependent on the temperature $T$, it becomes necessary to stabilize the $G_{\mathrm{APD}}$ in real time by adjusting the bias voltage $V_{\mathrm{bias}}$ applied to the APDs in response to $T$ changes \cite{alimenti2024design,alimenti2024development}.  
To enable this, the payload must include high-voltage (HV) sources capable of performing such regulation.
In order to achieve this, we selected precision-regulated high-voltage DC/DC converters from the SMHV series by HVM Technology Inc., featuring a sub-miniature case size of $0.85''\times0.85''\times0.60''$. These compact modules are well suited for CubeSat missions due to their small footprint and efficient performance. A series of tests is planned to evaluate the performance of the HV modules. The main objective is to characterize the high-voltage output by verifying the stability of the output voltage under variations in the input supply, assessing the effectiveness of the voltage regulation and current-limiting features, and analyzing the output ripple. The ripple characterization will include measurements of both the peak-to-peak voltage and the frequency components, using a digital oscilloscope to ensure a precise and detailed analysis of the output signal. Thus, in this work, we show and discuss the measurement system that was specifically designed for this aim.

\section{Measuring systems design}

In this section, the key parameters that we intend to investigate during the preliminary tests are presented, along with a description of the dedicated measurement system. Given the strong temperature dependence of the avalanche photodiode (APD) gain \cite{alimenti2024design}, it is crucial to apply a bias voltage that is dynamically adjusted based on the APD temperature, in order to ensure stable and reliable operation. Prior to the integration of the SMHV DC/DC converters, a comprehensive set of preliminary tests is thus required. These tests are designed to evaluate the performance, safety, and reliability of the converters under conditions representative of the final application environment.

\subsection{Definition of test objectives and parameters for power module validation}\label{sec:Tests}

To ensure the suitability of SMHV DC/DC converters (\figurename~\ref{fig:enter-label}) for powering sensitive photodetectors such as APDs and MAPMTs, a dedicated measurement system is under development. This system will be designed to conduct a comprehensive set of electrical and functional tests aimed at validating key performance aspects of the converters under representative operating conditions. The proposed test framework will enable the following characterizations:

\begin{figure}[ht]
  \centering
     \includegraphics[width=0.40\linewidth]{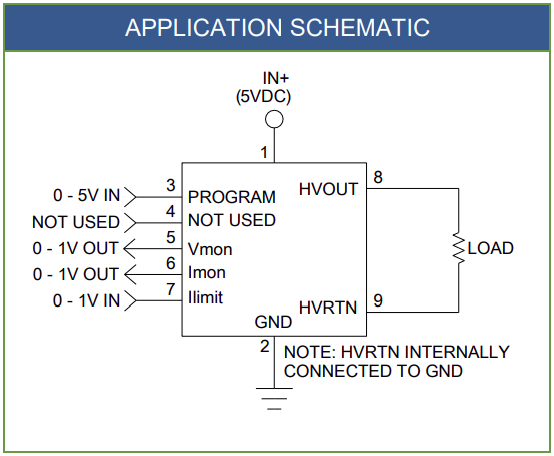}
     \caption{SMHV0505 pin-out.}
     \label{fig:enter-label}
\end{figure}

\begin{itemize}
\item {Programmable output control characterization:}  
   the converter's output voltage $V_\mathrm{out}$ programmability will be characterized by sweeping the control input $V_\mathrm{ctrl}$ (pin PROGRAM in \figurename~\ref{fig:enter-label})  across its full operational range $[0\;\mathrm{V};5\;\mathrm{V}]$ and measuring the corresponding output voltage. This test will allow accurate determination of the transfer function $ {V_\mathrm{out} = f(V_\mathrm{ctrl})} $, including linearity, resolution, gain, repeatability, and potential hysteresis. Such characterization is essential to ensure precise and predictable output voltage setting during operation, particularly in applications requiring dynamic controls.

  \item {Output voltage regulation and line stability:}  
  the test setup shall evaluate the line regulation performance of the converter by introducing controlled variations to the input voltage and measuring the resulting changes in the high-voltage (HV) output. The aim is to characterize the converter ability to maintain a stable output under typical power supply fluctuations, as expected in the target application environment.

  \item {Overcurrent protection characterization:}  
  to assess the functionality of the internal current limiting circuitry, the converter will be set with different limits of the output current and subjected to controlled overcurrent scenarios, such as short-duration low-impedance loads. The test will determine the current threshold at which the protection engages, the nature of the limitation, and the system’s ability to recover normal operation once the fault is cleared.

  \item {Output ripple and noise analysis:}  
  the high-frequency noise and ripple superimposed on the DC output will be measured using appropriate filtering and bandwidth-limited instrumentation. The analysis will include the root mean square (RMS) and peak-to-peak voltage of the ripple, as well as spectral components, to evaluate their potential impact on analog front-end noise performance in the photodetectors readout circuits.

  \item {Load regulation and static load response:}  
  the converter’s performance under different steady-state loading conditions will be assessed using precision resistive loads. This test is intended to verify output voltage accuracy across the load range, providing a baseline characterization before proceeding with tests involving active or sensitive electronic subsystems.

\item {Temperature-dependent performance evaluation:}  
  to assess the thermal stability of the converter, a comprehensive set of tests will be conducted in a controlled thermal environment, such as a thermal chamber, across the expected temperature range of the space mission. Key performance parameters—including output voltage, line and load regulation, ripple, and control input response—will be measured at various setpoints from $-20\,^{\circ}\mathrm{C}$ to $+60\,^{\circ}\mathrm{C}$. This analysis aims to identify temperature-induced drifts or anomalies and to validate the converter’s suitability for operation under orbital thermal cycling conditions.

\end{itemize}



The implementation of this test methodology represents an important step toward verifying the performance of the power modules prior to their integration. This process helps reduce the likelihood of malfunction and supports the overall reliability of the system.

\subsection{Experimental set-up}

\begin{figure}[ht]
        \centering
        \includegraphics[width=0.9\linewidth]{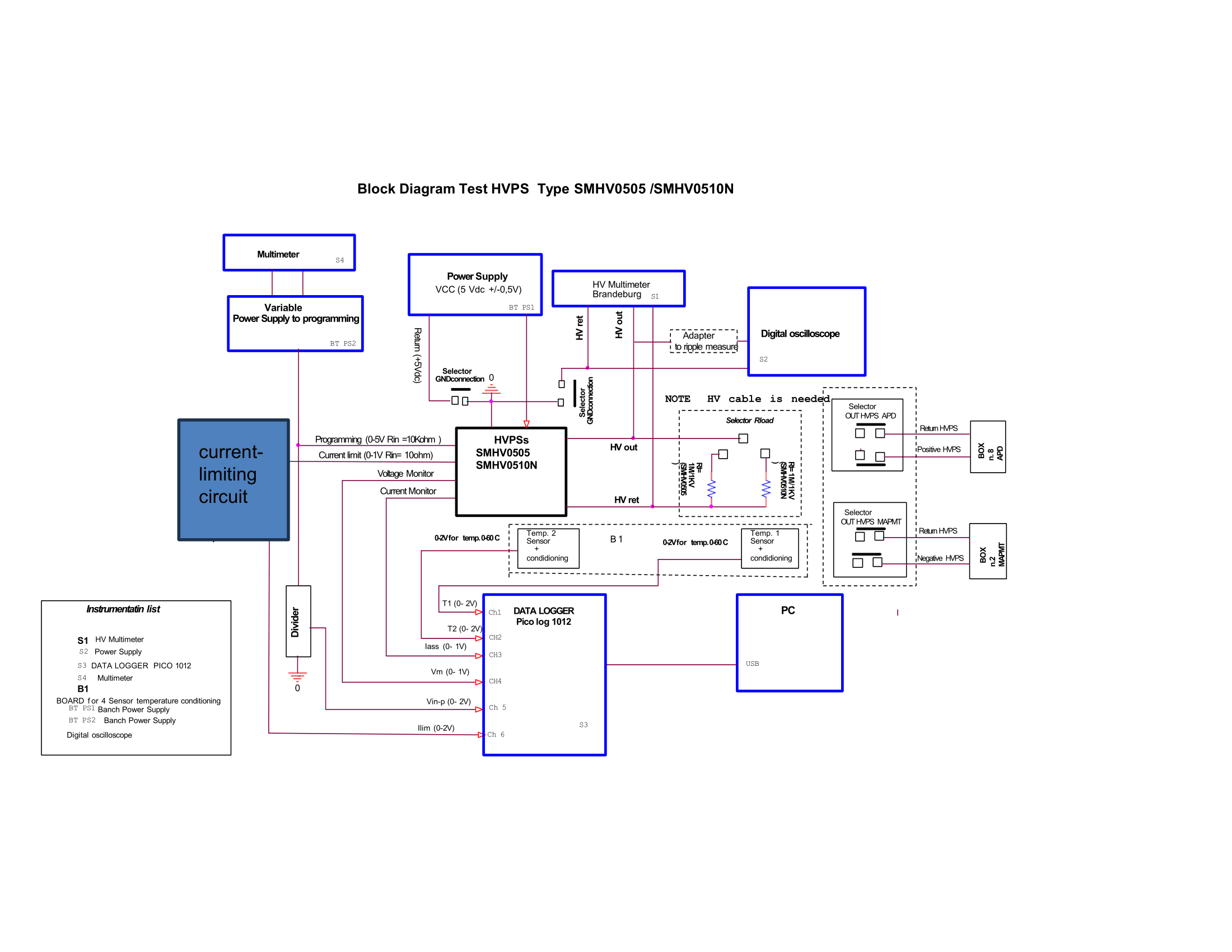}
        \caption{Block diagram of the test set-up}
    \label{fig:BlockDiagram}
\end{figure}

The experimental setup designed to perform functional tests of the SMHV DC/DC converters is illustrated in the block diagram shown in \figurename~\ref{fig:BlockDiagram}. This setup was specifically designed to enable precise and reproducible measurement of all the relevant electrical parameters discussed in Sec.~\ref{sec:Tests} while ensuring safety during high-voltage testing.

The block diagram outlines the architecture of the experimental test bench designed for the characterization of the SMHV DC/DC converters. The system comprises the HV power module, a programmable control interface, precision measurement instrumentation, thermal monitoring elements, safety circuitry, and configurable resistive loads. The setup supports real-time acquisition of key parameters—including output voltage, ripple, current limiting behavior, and thermal dependence—under controlled test conditions. The following items detail the primary functional blocks of the setup:

\begin{itemize}

  \item Programmable control interface:  
  a low-voltage power supply is used to program the HV output via the converter’s control input pin. The programming voltage is monitored with an accurate digital multimeter. 

  \item Primary power input:  
  the converter is supplied by a regulated low-voltage DC source (typically 5~V~±~0.5~V). Low ripple and noise levels on this supply are essential to ensure proper converter operation and to avoid spurious effects on output stability. This power supply, eventually substituted by a Source Meter Unit (SMU), will also allow measuring the power consumption and conversion efficiency for different loads and output voltages.

\item Current limit programming interface:  
the DC/DC converter includes an internal current limiting circuit whose threshold is adjustable via an analog control pin. The test bench incorporates a programmable source capable of sweeping the control voltage from 0~V to 1~V, corresponding linearly to 0\% to 100\% of the maximum output current. This programmable current limit interface enables characterization of the converter overcurrent response and protection behavior under controlled and reproducible conditions.

  \item High-voltage measurement unit (HV multimeter):  
  high-voltage output measurements are performed using an HV multimeter, which features high input impedance and precision scaling, enabling accurate and safe measurement of voltages in the kilovolt range.

  \item Oscilloscope for ripple analysis:  
  a digital oscilloscope is used to measure high-frequency ripple and transient behavior on the HV output. Time-domain measurements of peak-to-peak ripple amplitude and spectral content are essential for assessing the converter noise performance.

  \item Thermal monitoring:  
  two temperature sensors are positioned to monitor both ambient and local thermal conditions in proximity to the converter. These readings enable correlation between output variations and temperature, facilitating compensation strategies.

  \item Data acquisition system (PicoLog 1012):  
  the PicoLog 1012 DAQ unit is used to acquire multiple analog signals, including programming voltage, $V\textsubscript{mon}$, $I\textsubscript{mon}$, temperature, and current limit feedback. Continuous acquisition and USB interfacing with a PC allow for real-time monitoring and long-term logging.

\end{itemize}

\section{Summary}

This paper outlines the methodologies, instrumentation, and system architecture developed to perform comprehensive testing of the SMHV series high-voltage DC/DC converter. The test bench is designed to enable precise control, detailed monitoring, and accurate regulation of the output voltage, aiming to evaluate stability, safety, and compliance with the stringent requirements of spaceborne applications. The setup integrates temperature sensors to monitor thermal variations, whereas accurate measurement instruments, including a high-voltage digital multimeter, a digital oscilloscope, and a PicoLog 1012 data acquisition system, allow real-time monitoring of voltage, current, temperature, and output ripple. Signal conditioning circuits are employed to ensure proper interfacing between sensors and acquisition hardware. The proposed test architecture will facilitate the experimental characterization of the HV converters under representative electrical and environmental stresses encountered in orbit. Continuous data acquisition will support in-depth post-processing analysis, contributing to the validation of the module operational reliability.

\acknowledgments 
This work is funded by the Italian Space Agency (ASI) within the Alcor Program, as part of the development
of the CUbesat Solar Polarimeter (CUSP) mission under ASI-INAF contract n. 2023-2-R.0.

\bibliography{report} 
\bibliographystyle{spiebib} 

\end{document}